\documentclass[twocolumn]{aastex631}

\usepackage{enumerate}
\definecolor{forestgreen(web)}{rgb}{0.13, 0.55, 0.13}
\definecolor{ao}{rgb}{0.0, 0.0, 1.0}
\definecolor{lightbrown}{rgb}{0.71, 0.4, 0.11}

\hypersetup{linkcolor=red,filecolor=cyan,urlcolor=ao}

\def\be{\begin{equation}}
\def\ee{\end{equation}}
\def\ba{\begin{eqnarray}}
\def\ea{\end{eqnarray}}

\usepackage{multirow}

\shorttitle{Cosmic rays in Milky Way CGM}
\shortauthors{Roy, Nath}

\begin{document}

\title{Constraints on cosmic rays in the Milky Way circumgalactic medium from OVIII observations}
\correspondingauthor{Manami Roy}
\email{manamiroy@rri.res.in}
\author[0000-0001-9567-8807]{Manami Roy}
\author[0000-0003-1922-9406]{Biman B. Nath}
\affiliation{Raman Research Institute, 
Sadashiva Nagar, 
Bangalore 560080, India\\ {\rm Accepted for publication in ApJ on Apr 22, 2022}}

\begin{abstract}
We constrain the cosmic ray (CR) population in the circumgalactic medium (CGM) of Milky Way by comparing the {observations} of absorption lines of OVIII ion 
 {with predictions from analytical models of CGM {: precipitation (PP) and isothermal (IT) model.} For a CGM in hydrostatic equilibrium, the introduction of CR suppresses {thermal pressure}, and affects the OVIII ion
 abundance. We explore the allowances given to the ratio of CR pressure to thermal pressure ($\rm{P}_{\rm{CR}}/\rm{P}_{\rm{th}}=\eta$), with varying boundary conditions, CGM mass content, photoionization by extragalactic ultraviolet background and temperature fluctuations. We find that the allowed maximum values of $\eta$ are : {$\eta\lesssim10$ in the PP model and $\eta\lesssim6$ in the IT model. {We also explore the  spatial variation of $\eta$ : rising ($\eta=A x$) or declining ($\eta=A/x$) with radius, where A is the normalization of the profiles}. In particular, the models with declining ratio of CR to thermal pressure fare better than those with rising ratio {with} suitable temperature fluctuation (larger $\sigma_{\rm ln T}$ for PP and lower for IT). The declining profiles allow $A\lesssim8$ and $A\lesssim10$ in the case of IT and PP models, respectively, {thereby accommodating} a large value of $\eta \,(\simeq 200)$ in the central region, but not in the outer regions.}  These limits, combined with the limits derived from $\gamma$-ray and radio background, can be useful for building models of Milky Way CGM including CR population. {However, the larger amount of CR can be packed in cold phase which may be one way to circumvent these constraints.}}
\end{abstract}


\section{Introduction}
Galaxies have two components:  galactic disk  surrounded by a gaseous and a dark matter halo. The gaseous halo, which extends up to the virial radius (sometimes even beyond) is known as the circumgalactic medium (CGM). It is a reservoir for  most of the baryons and plays a crucial role in galaxy formation and evolution by various feedback processes such as outflowing, recycling of gas \citep{Tumlinson2017}.  Soft $X$-ray observation of OVII and OVIII absorption \citep{Gupta2012, Fang2015} and emission lines \citep{Henley2010, Henley2_2010}, and SZ effect \citep{Planck2013,Anderson2015} indicate the presence of a hot phase (T $\geq 10^6$ K) of CGM. Recently, a warm  ($10^5 \rm{K}<T<10^6$K) and a cool phase ($10^4 \rm{K}<T<10^5$K) of CGM have also been discovered through absorption lines of low  and intermediate ions  \footnote{Low Ions: Ionization Potential (IP) $< 40$ eV, T $= 10^{4-4.5}$ K; Intermediate Ions: $40 \gtrsim$ IP (eV) $\lesssim 100$, T $= 10^{4.5-5.5}$ K; {High Ions: IP$ \gtrsim 100$, T$>10^{5.5}$K.}} \citep{Tumlinson2017}    at low redshifts \citep{Stocke2013,Werk2014,Werk2016,Prochaska2017} and Ly$\alpha$ emission at high redshifts \citep{Hen2015,Cai2017}. 
It is not yet clear how this cool phase coexists with the hot phase and survives the destructive effects of various instabilities \citep{McCourt2015,Ji2018}, but this discovery has led to a picture of multiphase temperature and density structure of the CGM \citep{Tumlinson2017,Zhang2018}. It also partially solves the problem of missing baryon in the galaxies \citep{Tumlinson2017}. 

\begin{figure}
\includegraphics[width=0.5\textwidth]{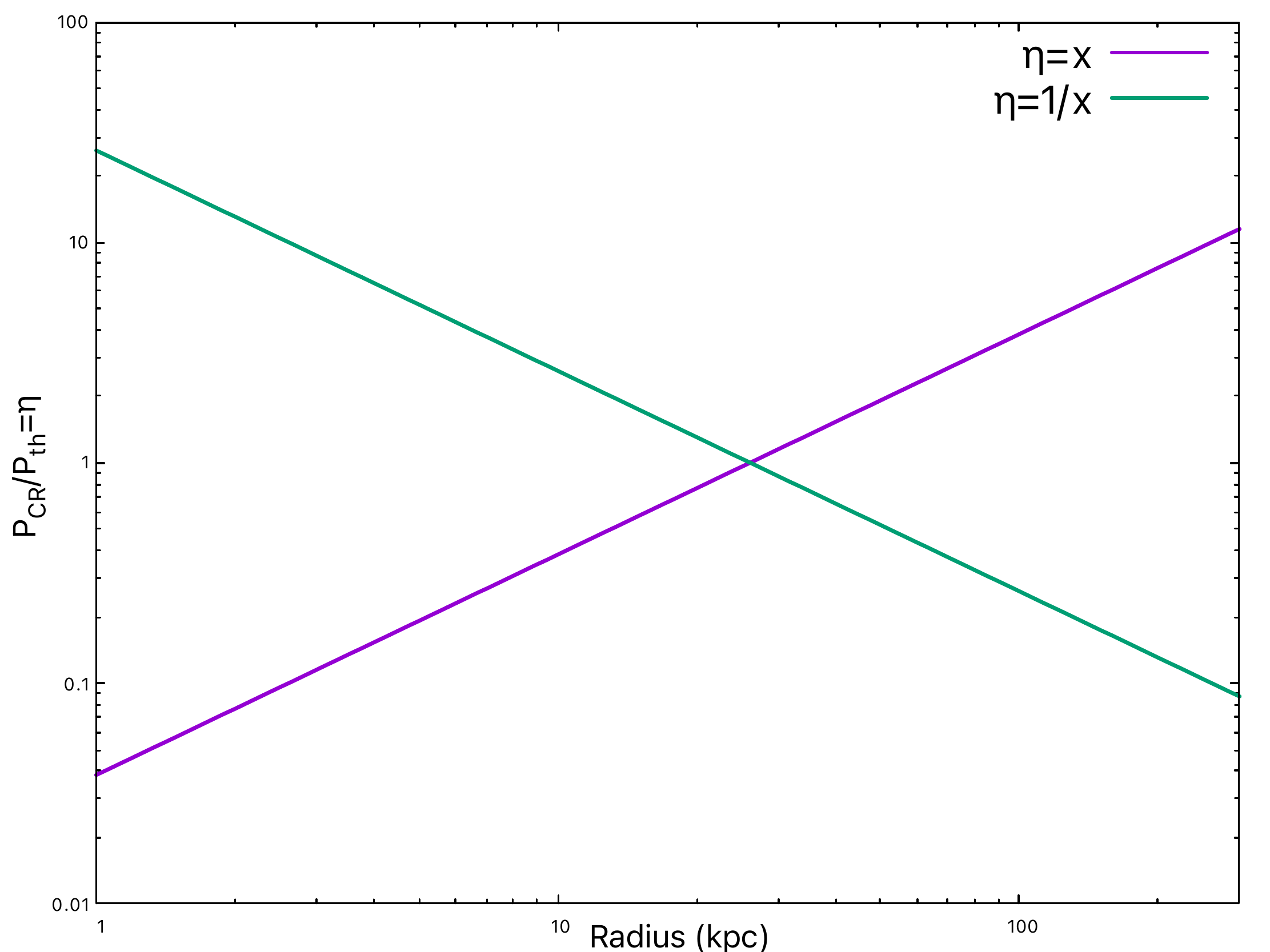}
\caption{{Radial profiles of $\eta$ in the case of $\eta=$x and 1/x.}}
\label{eta_profiles}
\end{figure}

{If the CGM is considered to be in hydrostatic equilibrium by means of only thermal pressure, then in order to maintain pressure equilibrium the cold component ($\sim 10^4$ K) of CGM is expected to have a higher density than the hot component ($\sim 10^6$ K).}   
However, \cite{Werk2014} found the density of cold gas to follow the hot gas density distribution. This has given rise to the idea of a non-thermal pressure component, consisting of cosmic-ray (CR) and magnetic pressure which can add to the thermal pressure. 
Without a non-thermal component, the abundances of low and mid-ions (seen in cool and warm phases, respectively) are under-estimated even in  high-resolution simulations \citep{Voort2019,Hum2019,Pee2019}.
\cite{Ji2019} have suggested that 
CR can 
explain {these two problems}. 
It has been suggested that CR can provide pressure support to cool diffuse gas \citep{Salem2015,Butsky2018}, help in driving galactic outflows \citep{Rus2017,Wiener2017}, as well as excite 
Alfv\'en waves that {can heat}
the CGM gas \citep{Wiener2013}. 

The magnitude of the CR population in the CGM is, {however}, highly debated. Some simulations \citep{Butsky2018,Ji2019,Dashyan2020} claimed a CR dominated halo, which increases feedback efficiency of the outflowing gas by increasing the mass loading factor and suppressing the star formation rate. The ratio of CR pressure and thermal pressure ($\rm{P}_{\rm{CR}}/\rm{P}_{\rm{th}}=\eta$) controls this effect and can reach a large value exceeding 10 over a huge portion of Milky-way sized halo \citep{Butsky2018}. Another simulation claimed that MW-sized halo at low redshift (z$<1$) can be CR populated with $\eta$ nearly $10$ in outflow regions, although in warm regions (T=$10^5$ K) of halo, $\eta\leq1$ \citep{Ji2019}. { \cite{Dashyan2020} found in their simulation that dwarf galaxies, with virial mass $10^{10}-10^{11}$ M$_\odot$, can have a high value of $\eta$ ($\sim100$) at $1-5$ kpc from the mid-plane when isotropic diffusion is incorporated (their figure 1 and 2 ).} 

It is therefore an important question as to  how much CR can be accommodated in the CGM in light of different observations. 
One of the most obvious effects of CR population in CGM is to suppress the thermal pressure of the hot phase
which is also seen in the simulation of \cite{Ji2019}. The recent work by \cite{Kim2021} pointed out that this
reduction in thermal pressure would lower the thermal SZ (tSZ) signal, as tSZ probes the integrated thermal
pressure in the halo. This diminished value of thermal pressure would also significantly change the abundance of
high ionization species, and thereby jeopardise the interpretation of their column densities. In a recent work by
\cite{Faerman2021}, they  modified their previous isentropic model \cite{Faerman2019} with significant
non-thermal pressure ($\alpha=(P_{nth}/P_{th}) +1.0=2.9$) and found that the gas temperature  in the central
region for the non-thermal model becomes lower than the value at which OVIII CIE temperature peaks, which in turn results in a lower OVIII column density. At low CGM
masses, photoionization compensated for this effect by the formation of OVIII at larger radii, but for large gas masses and large mean densities (as in the Milky Way), photoionization effect is negligible and the total OVIII column density is low. In a previous work \citep{Jana2020}, we constrained the CR pressure ($\eta=P_{CR}/P_{th}$) in the CGM in light of the 
isotropic $\gamma$-ray background (IGRB) and radio background,  using different analytical models (Isothermal Model: IT, Precipitation Model: PP) of CGM. In the case of IT model, the value of IGRB flux puts an upper limit of 3 {on $\eta$, whereas all values of $\eta$ are ruled out if one considers the anisotropy of the flux due to off-center position of solar system in Milky Way.} However, in the case of PP model, IGRB flux value and its anisotropy allow a
range of $\eta$ from 100 to 230. In comparison, the constraints from radio background are not quite robust. 
In this paper, we use the same analytical models (IT and PP) and   
put limits on CR population in the CGM of Milky Way by comparing the predicted OVIII absorption column densities (N$_{\rm OVIII}$) with the available observations. 

It might still be asked, why use OVIII as a probe? Previous works \citep{Faerman2017,MR2021} explained the observed OVII, OVIII absorption column densities in the Milky Way without the incorporation of CR component using the analytical models of CGM used here. It is, therefore, important to study the effect of CR on these column densities with the inclusion of CR component in these models. Whereas the OVII ion has a plateau of favourable temperatures ($5\times10^5$ K - $1\times10^6$ K),  the suitable temperature for OVIII production ($2\times10^6$K) peaks near the temperature of the hot halo gas of Milky Way CGM. This particular fact makes OVIII abundance a  sensitive probe of the CGM hot gas and motivates the choice of OVIII column density for comparison with the observations in order to constrain CR population in Milky Way CGM.


\section{Density and Temperature Models}
\label{profile}
 We study two {widely discussed} analytical models: 1) Isothermal Model (IT), 2) Precipitation Model (PP). {These models have previously been studied {in the context of} absorption and emission lines from different ions
 \citep{Faerman2017,Voit2019,MR2021}} We assume the CGM {to be} in hydrostatic equilibrium {within} the potential of dark matter halo {($\phi_{\rm DM}$)} {, with a metallicity of} $0.3$ Z$_\odot$ \citep{Pro2017}, {so that,} 
 {
 \begin{equation} \label{HSE1}
    \frac{dP_{\rm total}}{dr}  = -\frac{d\phi_{\rm DM}}{dr}\rho, 
\end{equation}}
{where $P_{\rm total}$ is the total pressure and $\rho$ and r are the density and radius, respectively. We consider the Navarro-Frenk-White (NFW) \citep{NFW} profile as the underlying 
dark matter potential in the IT model. 
However we slightly modify this 
potential in the case of the PP model as 
suggested in \cite{Voit2019} 
by considering {the} circular velocity (v$_c$) to be constant (v$_{\rm c,max}=220$ km s$^{-1}$) up to a radius of $2.163$ {$\rm r_{\rm vir}/\rm c$} for a halo with M$_{\rm vir}=$ $2\times10^{12}$ M$_\odot$ and concentration parameter $c=10$.} 
We include non-thermal pressure support with CR population and magnetic field {along with thermal pressure} in these models. {We consider the magnetic energy to be {in} equipartition with thermal energy (P$_{\rm{mag}}$=$0.5$P$_{\rm{th}}$) so that {the} total pressure $\rm{P}_{\rm total}=\rm{P}_{\rm{th}}(1.5+\eta)$, which leads to,} 
 {
 \begin{equation} \label{HSE2}
        \frac{dT}{dr} = - \left( \frac{1} {1.5+\eta} \right) \left( \frac{\mu m_p} {k} \right)\frac{d\phi_{\rm DM}}{dr}  
            - \left( \frac{T}{n} \right)  \frac{dn}{dr} \,.  
\end{equation}}
 {{Note that the magnitude of magnetic field in CGM is rather uncertain. While \cite{Bernet2008} claimed the magnetic field in CGM of galaxies at $z=1.2$  to be larger than that in present day galaxies, the observations of \cite{Prochaska2019} suggested a value less than that given by equipartition. Yet another study claimed near-equipartition value after correlating the rotation measures of high-$z$ radio sources with  those in CGM of foreground galaxies \citep{Lan2020}.}}
 
 {The inclusion of non-thermal pressure} {suppresses} the thermal pressure{, and} in turn the temperature or density or both of the CGM gas (also seen in recent simulation of \cite{Ji2019}). Our goal is to determine the effect of {the inclusion of} CR population on N$_{\rm OVIII}$ therefore constrain the CR population in Milky Way CGM. {In addition to models with uniform $\eta$, we have studied the effect of varying $\eta$ as a function of radius. For this, we have explored two contrasting variations, $\eta=A \times x$ (rising profile) and $\eta= A/x$ (declining profile) where x=r/r$_{\rm s}$, where A is normalization of the profiles. We primarily studied the case with A=1, where the maximum value (for $\eta=x$) is r$_{vir}/r_{s}$=c=10. For $\eta=1/x$, to avoid divergence at r=0 kpc we have started our calculation from r=1 kpc, therefore the maximum value in this case is r$_s/1$ kpc$=26$. {We have shown the profiles of $\eta=x,1/x$ in the figure \ref{eta_profiles}.} We have explored the {effects of scaling up and down the values with these profiles} as well {(e.g., $\eta=Ax, B/x$, with $A,B\ge 1$)}.  
 Although $\eta$ can vary in a more complicated manner, we study these two particular cases (increasing and decreasing linearly) as the simplest representatives of a class of models in which $\eta$ is allowed to vary with radius.} {It should be noted that the $\eta$ here denotes the CR pressure in the diffuse hot gas with respect to thermal pressure of the hot gas.} 

 \begin{figure*}
\includegraphics[width=\textwidth]{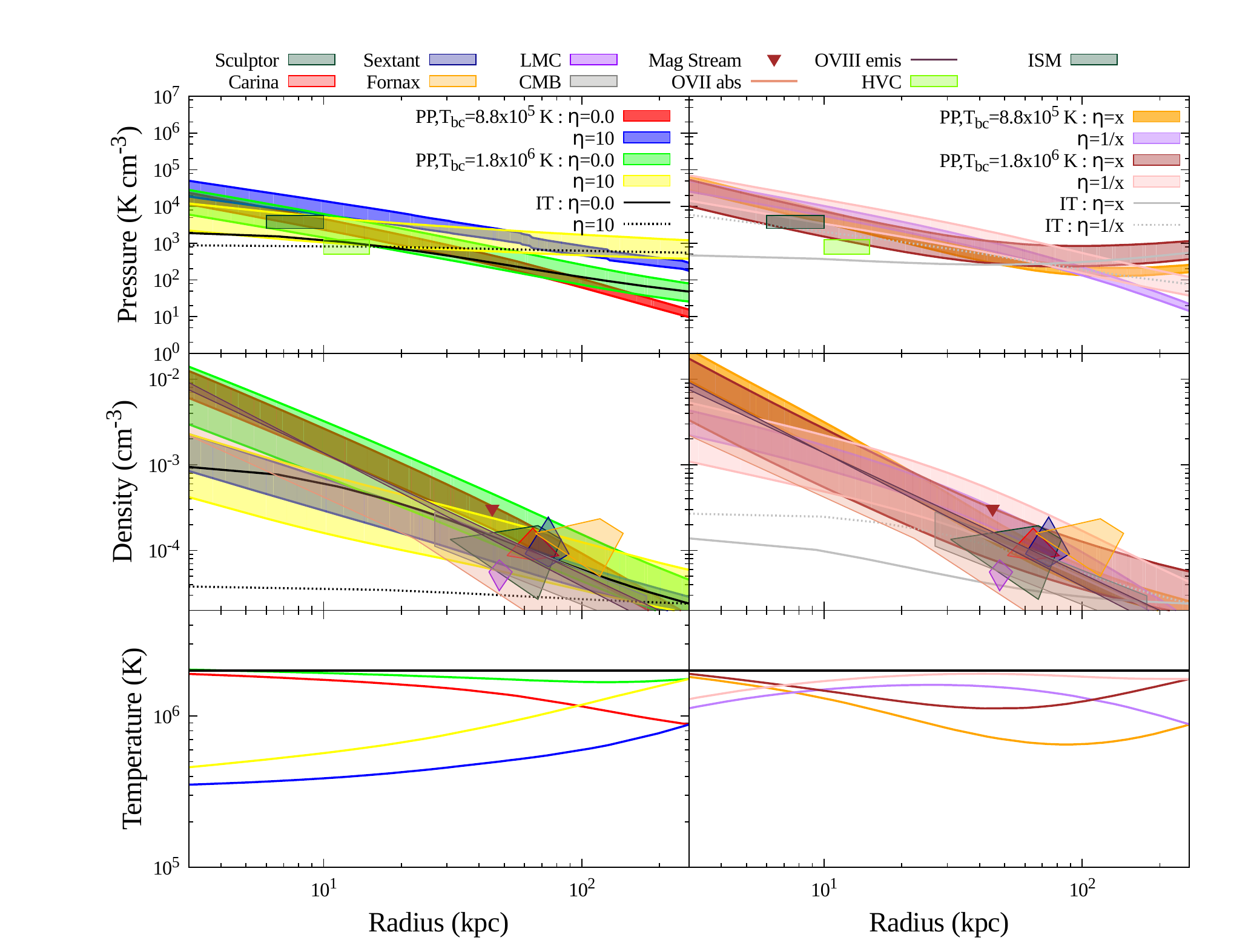}
\caption{{Total pressure (top panel), density(middle panel) and temperature (bottom panel) profiles for different models (IT and PP model)  without ($\eta=0$), with constant CR population ($\eta=10)$ and CR profile ($\eta=$x and 1/x), where x=r/r$_{\rm s}$. Different observational constraints of pressure and density are also shown.}}
\label{profiles}
\end{figure*}


{There are several ways to proceed, starting from equation \ref{HSE2}. One way is to simply consider the temperature all over the CGM to be constant, which implies the LHS of the equation \ref{HSE2} to be zero. Another way is to use  specific entropy to relate the two unknown quantities in the equation \ref{HSE2}, temperature and density. We discuss these two ways in details in the following subsections. }
 

\subsection{Isothermal Model}
The IT model {provides the simplest model for the} CGM gas {with a} constant {temperature}, so that,
{
 \begin{equation} \label{IT}
        \frac{dn}{dr} = - \left( \frac{1} {1.5+\eta} \right) \left( \frac{\mu m_p} {k\,T} \right)\frac{d\phi_{\rm DM}}{dr}\,\times n  \,.
\end{equation}}
Previously hot gas has been modelled using isothermal profile by \cite{MB2015} where they fitted their model with the OVIII emission line observations which determined the hot gas mass to be within the range of $\sim 2.7-4.7\times10^{10} M_\odot$. Therefore, we normalize the density profile isothermal model without any non-thermal component such that the hot gas mass is $5\times10^{10} M_\odot$. {It should be noted that the single temperature IT model does not represent the  multiphase nature of the CGM. However, OVIII abundance is sensitive mostly to the hot phase of CGM which is well described by the single phase model. Therefore, our limits from the IT model will not {change} even if we have a two or three temperature model. Another way of incorporating multi-phase in IT model is to consider temperature distribution around the single mean temperature. {We will show} below that incorporation of such distribution in IT model will tighten our limit on CR. We can, {therefore,} treat the limit from our single temperature IT model as an upper limit.} We then include a non-thermal pressure component in our model along with the thermal pressure. At this point, there are {again} several ways to proceed. One can fix the mass, and hence the density profile of the hot gas, and compensate for the decrease in thermal pressure by reducing the temperature \citep{Jana2020}. Another way is to fix the temperature and reduce the density by keeping the boundary value same. However, {for} massive galaxies (M$_{\rm vir} \geq 10^{12}$ M$_\odot$) \citep{Li2015} as well as Milky Way \citep{MB2015}, the {halo} temperature  is observed to be $\geq 2 \times 10^6$ K. 
This motivates us to assume a constant CGM temperature of $2\times10^6$ K  for all the IT model (without and with non-thermal pressure) \citep{MB2015}.  We therefore vary the density profile for different values of $\eta$ by keeping the density at the outer boundary fixed. This lowers the inner density for increasing value of $\eta$, and for $\eta=10$, the hot gas mass becomes $2.7 \times 10^{10} M_\odot$ - the lower limit of hot gas mass obtained by \cite{MB2015}. We, therefore, do not consider larger values of $\eta\ge 10$ for IT model. In the left panel of Figure \ref{profiles}, density profiles for IT model are shown by black solid ($\eta=0$) and 
dotted lines ($\eta=10$). The grey solid and dotted lines in the right panel of Figure \ref{profiles} denote the $\eta=$x and 1/x cases respectively.

\begin{figure}
\includegraphics[width=0.5\textwidth]{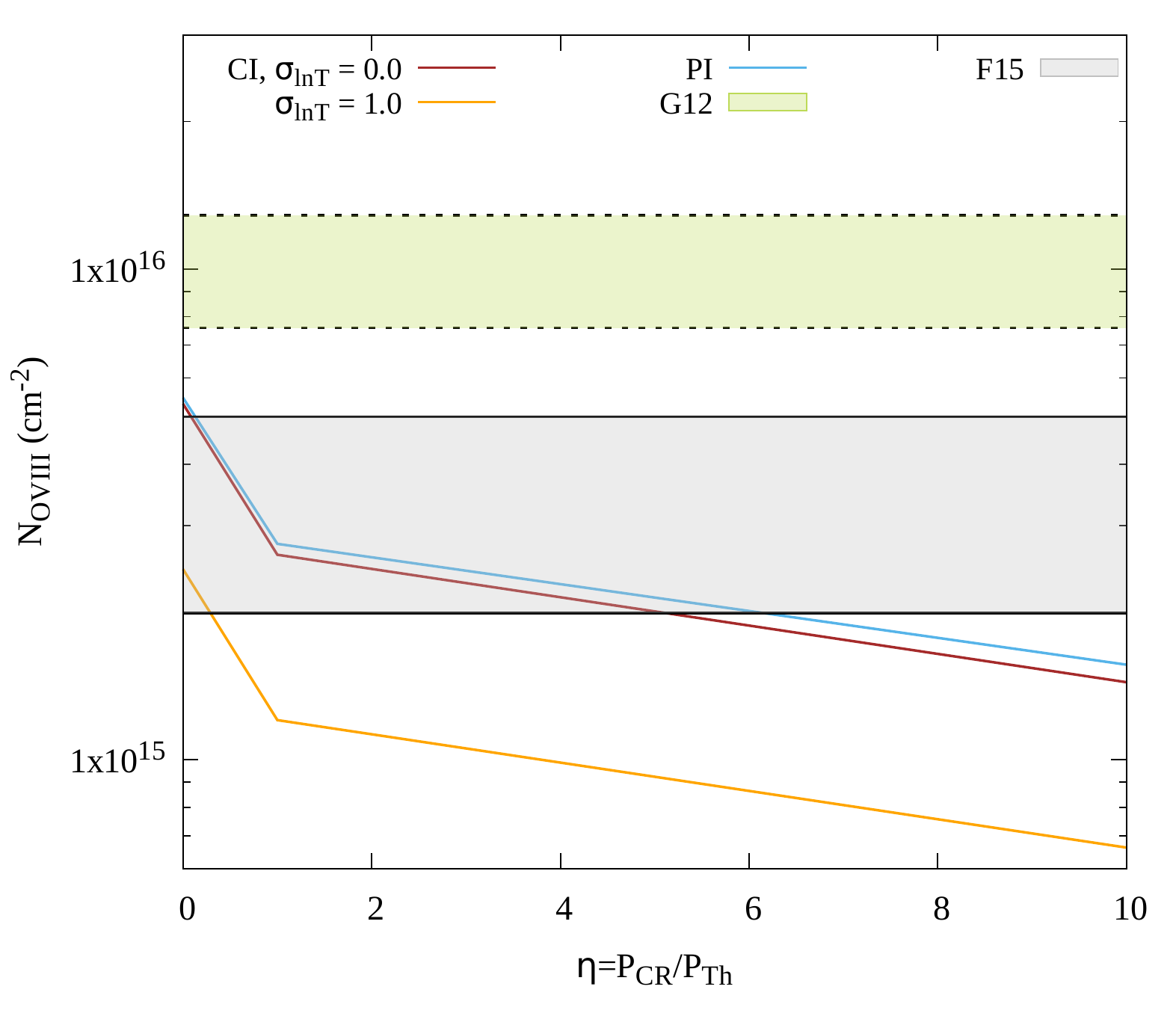}
\caption{{Variation of OVIII column density with $\eta$ for IT model with different cases like CI without fluctuation (brown),  with fluctuation (orange) and inclusion of PI (blue) with observational constraints by shaded regions from G12 (green) and F15 (grey). }}
\label{OVIII_IT}
\end{figure}

\subsection{Precipitation Model}
{The PP model is a physically motivated CGM model which uses the specific entropy as a connection between density and temperature. It is built upon the concept of a threshold limit for the ratio of radiative cooling time (t$_{\rm cool}$) to free fall time (t$_{\rm ff}$). Recent phenomenological studies and numerical simulations {\citep[e.g.,][]{McCourt_2012MNRAS.419.3319M,Sharma2012,Voit2015}} showed that {thermally unstable perturbations can {lead to} multiphase condensation below this threshold value.} In this model, we consider a composite entropy profile ($K_{\text{pNFW}}(r)$) with the combination of base entropy and precipitation-limited entropy.}
{\begin{equation}\label{KpNFW}
  K_{\text{pNFW}}(r) = K_{\text{pre}}(r) + K_{\text{base}} (r)  \; ,
\end{equation}}
{where base profile is,} 
{\begin{equation} \label{Kbase}
    K_{\text{base}} = \left( 39 \, {\rm keV} \, {\rm cm}^2 \right) \,  
        v_{200}^2 \, \left( \frac {r} {r_{200}} \right)^{1.1} \; ,
\end{equation}}
{and precipitation-limited profile is,} 

\begin{equation} \label{Kpre}
    K_{\text{pre}}=\left[ \frac {2 kT(r)} {\mu m_p v_{\rm c}^2(r)} \right]^{1/3} 
        \left\{ \frac{10}{3} \left( \frac{2n_{\rm H}^2}{n_e n} \right)  
        \Lambda_{\text{N}} [ T(r)] \, r \right\}^{2/3} \; .
\end{equation}
Then using $n_e = ( kT / K_{\text{pNFW}})^{3/2}$, it follows from equation \ref{HSE2} that,
\begin{equation}
\frac{dT}{dx} \: = \: \frac{2}{5} 
    \left[ - \left( \frac{\mu m_H}{k} \right) \frac{d\phi}{dx} 
        + \frac{3}{2}\frac{T}{K_{pNFW}} \frac{d  K_{pNFW}}{d x} \right].  \, \\
     \label{PP}
\end{equation}
 {A recent study by \cite{Butsky2020} has shown the effect of cosmic ray pressure on thermal instability and precipitation for
different values of the ratio t$_{\rm cool}$/t$_{\rm ff}$. In their study, they found that large cosmic ray
pressure decreases the density contrast  of cool clouds.}  In {our} model, we keep t$_{\rm cool}$/t$_{\rm ff}$ {to be constant}  throughout the halo. Although observations and simulations of intercluster medium {point towards a value of this ratio} between $5-20$, it is  {a rather} uncertain parameter in case of CGM
studies and can have a wide range. We explore {a range of} the temperature boundary condition (T$_{bc}$) at {the} virial radius{, between} $\rm T_{\rm bc1}=0.5\mu m_{\rm{p}} {v_c}^2=1.8\times10^6$ K ($\simeq$virial temperature of the Milky-way) to $\rm T_{\rm bc2}=0.25 \mu m_{\rm{p}}{v_c}^2=8.8\times10^5$ K (used originally by \cite{Voit2019}). {The inclusion of non-thermal pressure in the model suppresses the temperature and in turn the density. However we keep the mass of the CGM gas constant by decreasing t$_{\rm cool}$/t$_{\rm ff}$ ratio. The variation of t$_{\rm cool}$/t$_{\rm ff}$ with  $\eta$ for different values of temperature boundary condition
can be seen in Figure \ref{profiles} of \cite{Jana2020}.
} We have taken the mass of the CGM in a range of $6\times10^{10}$ M$_\odot$ (from the original model of \cite{Voit2019} without CR) to $2.0\times10^{11}$ M$_\odot$ ({for} cosmic baryon fraction) in case of $T_{\rm bc1}$. However {the} mass range for $T_{\rm bc2}$ is $6\times10^{10}-1.0\times10^{11}$ M$_\odot$  {since beyond} this upper limit of mass{, the value of} t$_{\rm cool}$/t$_{\rm ff}$ falls below $1$.  We use {the} cooling function from CLOUDY \citep{Gary2017} for {a} metallicity of 0.3Z$_\odot$. 

\begin{figure*}
\includegraphics[width=\linewidth]{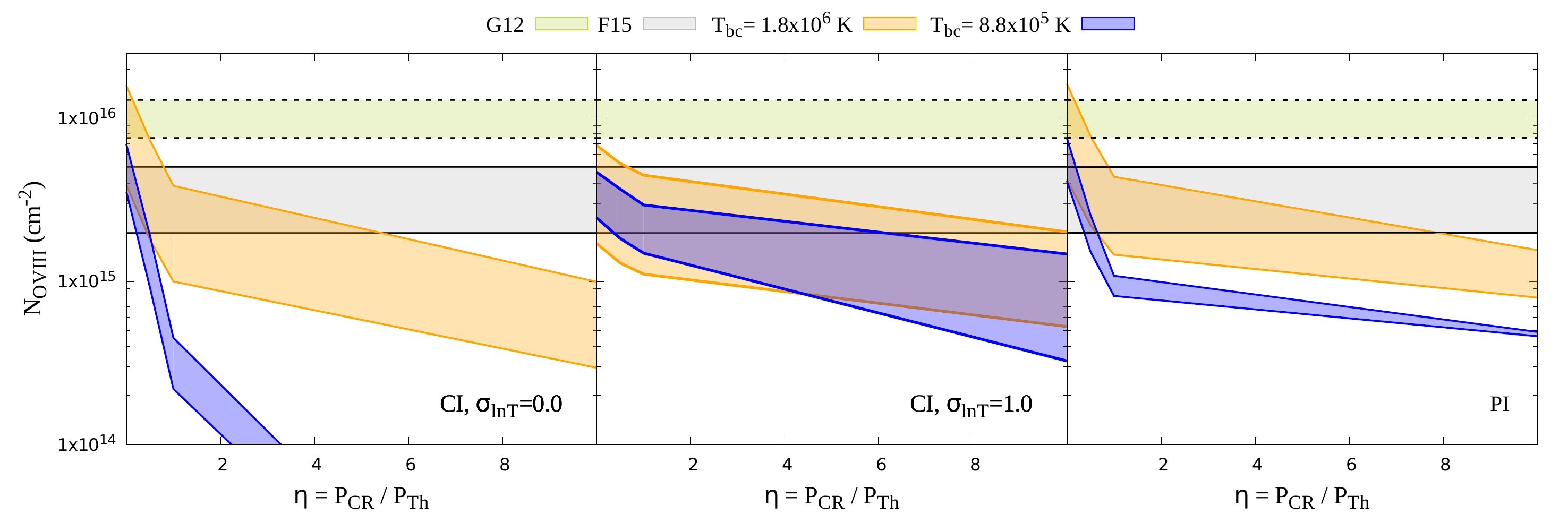}
\caption{{Variation of OVIII column density with $\eta$ for PP model with different cases such as CI without fluctuation (left panel), with fluctuation (middle panel) and with PI (right panel). The observational constraints are shown by shaded regions from G12 (green) and F15 (grey). Different temperature boundary conditions are shown by orange (T$_{bc}=1.8\times10^6$ K) and blue (T$_{bc}=8.8\times10^5$ K) shaded regions where shades denote the mass range for each case. }  }
\label{OVIII_PP}
\end{figure*}

{We show the temperature and density profiles for PP in Figure \ref{profiles} as shaded regions, bracketing the range of CGM mass mentioned above.} In {the left panel of} {Figure} \ref{profiles}, the temperature and density profiles are shown  {in} green and red for the PP model with no CR component whereas yellow and blue denotes $\eta=10$ for $\rm T_{\rm bc1}$ and $\rm T_{\rm bc2}$ respectively. {In {the right panel of} {Figure} \ref{profiles}, the temperature and density profiles are shown  {in} brown and orange  for the PP model with  $\eta=$x whereas pink and purple denote $\eta=$1/x for $\rm T_{\rm bc1}$ and $\rm T_{\rm bc2}$ respectively.}    

\subsection{Observational constraints}
Figure \ref{profiles} also shows observational limits on pressure, density from (a) observations of OVII and OVIII \citep{MB2015}, (b) CMB/X-ray stacking \citep{Singh2018}, (c) ram pressure stripping of LMC \citep{Salem2015}, Carina, Sextans \citep{Gatto2013},
Fornax, Sculptor \citep{Grcevich2009}, (d) high-velocity clouds \citep{Putman2012}, (e) Magellanic stream \citep{Stanimirovic2002}  and (f) ISM  pressure \citep{Jenkins2011}. We find that the obtained profiles including non-thermal components are within the observational limits. {We can also compare the temperature and density profiles with those from simulations that include CR. In the PP model, the distinguishing feature is a rising temperature profile from the inner to the outer region. CR simulations also show such profiles, e.g., as shown in Figure 5 of \cite{Ji2019}, especially beyond the galacto-centric radius of $\sim 10$ kpc. This particular profile from \cite{Ji2019} shows a decrease in temperature by a factor $\sim 7$ from the virial radius to $\sim 10$ kpc, similar to that shown in Figure \ref{profiles} for $\eta=10$ cases, although the outer boundary temperature is different in two cases. The difference in the profiles inward of $\sim 10$ kpc is not significant because the difference caused in the column density is small. The curves on the RHS of Figure \ref{profiles} for declining $\eta$ cases are similar to the simulation results expect at the outer radii, where PP models predict a declining temperature profile, which is not seen in CR-inclusive simulations. In brief, PP models with with constant or declining $\eta$ profiles appear to capture the CR-simulation temperature (and consequently) density profiles.}

\begin{figure}
\includegraphics[width=0.5\textwidth]{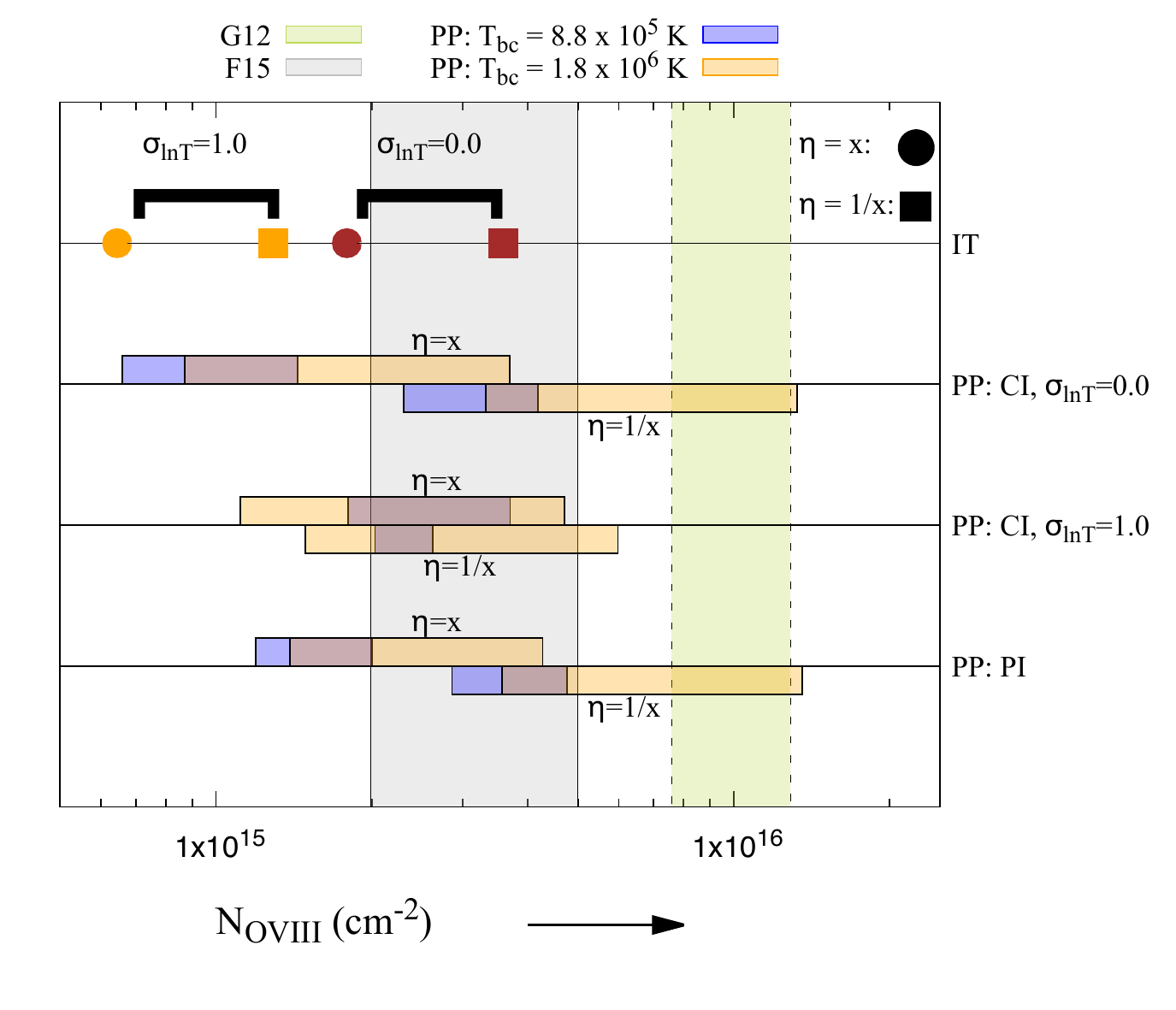}
\caption{{OVIII column density for different $\eta$ profiles for IT and PP model with different cases like CI without fluctuation,  with fluctuation and inclusion of PI with observational constraints by shaded regions from G12 (green) and F15 (grey). Different temperature boundary conditions are shown by orange (T$_{bc}=1.8\times10^6$ K) and blue (T$_{bc}=8.8\times10^5$ K) shaded regions where shaded region (above the lines for $\eta$=x and below the lines for $\eta$=1/x) denote the mass range for each case in PP model.  For IT model, brown and orange colours are used to denote $\sigma_{\rm lnT}=0.0$ and $1.0$ respectively whereas circles denote $\eta$=x and squares denote $\eta$=1/x.  }}
\label{OVIII_profile}
\end{figure}  
 
\subsection{Log-normal Temperature Fluctuation}
The observed widths and centroid offsets of OVI absorption lines in the CGM motivate us to consider dynamical disturbances causing temperature fluctuations in the CGM which results in a multiphase CGM gas. Low entropy gas parcels uplifted by outflows or high entropy gas can cool down adiabatically to maintain pressure balance and give rise to temperature and density fluctuations. Turbulence-driven nonlinear oscillations of gravity waves in a gravitationally stratified medium can be an alternative source for these fluctuations. Therefore, we consider {an} inhomogeneous CGM by including a log-normal temperature distribution {(characterized by} $\sigma_{\rm lnT}$) around the mean temperature.

{{Previous studies have} considered a log-normal temperature fluctuation for CGM and successfully {matched} with CGM observations. \cite{Faerman2017} considered log-normal distribution for {their} two-temperature ($1.5\times10^6$K and $5\times10^5$K) CGM model.} They found the best fit value of $\sigma_{\rm lnT}$ to be 0.3 in order to explain the observed OVI column density. {We have taken a similar approach. However we have a single temperature IT model. Therefore we consider a log-normal distribution of temperature around the single halo temperature in the case of IT model. In case of PP model, instead of a single temperature, we have a temperature profile $T(r)$. \cite{Voit2019,MR2021} took into account temperature fluctuation in PP model by considering log-normal distributions with constant value of $\sigma_{\rm lnT}$ around the mean temperature $T(r)$ at each radius.} \cite{Voit2019} found that $\sigma_{\rm lnT}=0.7$ satisfies the observed OVI column, whereas \cite{MR2021} concluded a range of $\sigma_{\rm lnT}=0.6-1.0$ in order to explain the observed OVII, OVIII and their ratio.  In the PP model, we follow the similar {approach} as in \cite{Voit2019,MR2021}.


\section{Results}
\label{column}
\begin{table*}
    \centering
    \caption{Constraints of $\eta$ on the basis of N$_{\rm OVIII}$ observations by G12 and F15}
 \begin{tabular}{c c c c c c c c} 
 \hline
 Model &  & \multicolumn{2}{c}{CI (without fluctuations)} & \multicolumn{2}{c}{CI (with fluctuations)} & \multicolumn{2}{c}{PI} \\ [0.5ex] 
 \hline & & G12 & F15 & G12 & F15 & G12 & F15 \\ 
 \hline\hline
 \multirow{3}{*}{PP}& T$_{\rm bc1}$ &  0.5 & 6 & -- & 10  & 0.5 & 8  \\ \\
 
  & T$_{\rm bc2}$ & --  & 0.5   & -- & 6  & -- & 1   \\ 
 \hline
 \hline
 \\
 IT & & -- & 5  & -- & 1  & -- & 6 \\ 
 \hline
\hline
 \end{tabular}
 \label{table}
 \end{table*} 

 We calculate the ionization fraction of OVIII using CLOUDY \citep{Gary2017}  {with the} input of density and temperature profiles derived from the CGM models. {{In order to incorporate} temperature fluctuation, we calculate the ionization fraction for all the temperatures in a log-normal distribution using CLOUDY and integrate them over the corresponding log-normal distribution to get a mean ionization fraction. That means, for IT model we get a single mean ionization fraction corresponding to the log-normal distribution around the single temperature of the halo. But for PP model, we will get mean ionization fractions at each radius corresponding to the log-normal distributions with constant value of $\sigma_{\rm lnT}$ around the mean temperature $T(r)$ at each radius.} We consider two cases: collisional ionization and photoionization along with collisional ionization. {We use} the extragalactic UV background \citep{HM12} at redshift $z=0$ for photoionization. We take {into account} our vantage point of observation i.e the solar position and calculate the variation of column density with the Galactic latitude and longitude. 
{We consider the median of the column densities for each value of $\eta$ in order to compare  with the observations.}

In {Figures} \ref{OVIII_IT} (for IT model) and \ref{OVIII_PP} (for PP model), we show the variation of N$_{\rm OVIII}$ with $\eta$, considering collisional ionization (CI) and photoionization {(PI)} for different model{s}. {One can  clearly see from these figures that with an increase in $\eta$, the value of N$_{\rm OVIII}$ decreases due to the decrease in the temperature.} In  {Figure}  \ref{OVIII_IT}, the {brown} and {blue} lines denote the cases with CI and PI respectively in the IT model. {In this figure, the orange colours denote the effect of log-normal fluctuations ($\sigma_{\rm lnT}=1.0$).} {The cases of CI and PI do not differ for {$\eta\le 1$}, {since} at $T\sim10^6$K, CI {dominates} over PI. However, {for $\eta\ge 1$}, PI {leads to} slightly larger {values of OVIII column density than the} CI case, as {further} decrease in temperature leads to more production of OVIII by PI. It should be noted that if one considers temperature fluctuation around a single favourable temperature of OVIII, there is less OVIII production with the increase in {$\sigma_{\rm lnT}$}.} 

{In {Figure} \ref{OVIII_PP} the orange shaded region indicates the boundary temperature near to virial temperature of halo $1.8\times10^6$ K (T$_{bc1}$) in PP model, whereas {the} blue shaded region denotes the boundary condition of $8.8\times10^5$ K (T$_{bc2}$) at virial radius. For the PP model, we show the cases with {CI without ($\sigma_{\rm lnT}=0.0$) and with fluctuation ($\sigma_{\rm lnT}=1.0$) and PI in} {left, middle and right panel of {Figure} \ref{OVIII_PP}}. The shaded regions in this figure refer {to} the {CGM} mass range {mentioned earlier }{in the case of PP model}. {{A boundary temperature that is close to the favourable temperature for OVIII yields more OVIII than one gets with lower boundary temperatures}. However, with the inclusion of temperature fluctuations, one gets almost similar OVIII for both cases. With lower boundary temperature, {the production of OVIII decreases for} $\eta\ge 2$, {because the inclusion of} $\eta$ {shifts the temperatures} from the CI peak of OVIII. However, {considering} PI in this case {can increase the amount of OVIII}.} }  

{In Figure \ref{OVIII_profile}, we show N$_{\rm OVIII}$ with different $\eta$ profiles for both of the models. For IT model, we use brown and orange colours to denote
$\sigma_{lnT} = 0.0$ and $1.0$ respectively whereas circles denote $\eta=$ x
and squares denote $\eta=$ 1/x. For PP model,  different temperature boundary conditions are shown by orange (T$_{bc}=1.8\times10^6$ K) and red (T$_{bc}=8.8\times10^5$ K) shaded regions where shaded region (above the lines for $\eta$=x and below the lines for $\eta$=1/x) denote the mass range for each case in PP model. {One {important} point to notice {here is} that declining $\eta$ profiles produce more OVIII than rising profiles.}}

\subsection{Observations To Compare With}
There are some soft X-ray observations of N$_{\rm OVIII}$ of Milky Way by \cite{Gupta2012}, \cite{MB2013} and \cite{Fang2015} {(Hereafter referred as G12, MB13 and F15 respectively)}. {G12} have measured N$_{\rm OVIII}$  along eight sight lines with \textit{Chandra} telescope. They have quoted the equivalent width (EW) of OVIII instead of N$_{\rm OVIII}$ in their paper. {However,} their measured values of EW of OVIII are increased by $30$\% for the correction of systematic error and column densities {were therefore re-calculated  by \cite{Faerman2017}. The re-calculated values of  log(N$_{\rm OVIII}$) are $16.0 (15.88-16.11)$ respectively. The {green shaded region} in {Figures} \ref{OVIII_IT}, \ref{OVIII_PP} and \ref{OVIII_profile} show the ranges of $N_{\rm OVIII}$ of {G12} as re-calculated by \cite{Faerman2017}. 

OVII lines have been studied by {MB13} with \textit{XMM-Newton} along the 26 sight-lines of distant AGNs as well as only one detection of OVIII with ratio of N$_{\rm{OVII}}$/N$_{\rm{OVIII}}$=$0.7 \pm 0.2$. {F15} observed the OVII absorption line for a broader sample of 43 AGNs which includes the sample from MB13. They reported a wider range of N$_{\rm OVII}$ ($10^{15.5-16.5}$ cm$^{-2}$) with the central value at $10^{16}$ cm$^{-2}$. However they have {arrived at these values} by {setting aside} the non-detection along 10 sight-lines. Adding these upper limits to the detected sample, \cite{Faerman2017} have come up with {a} new range of value of log(N$_{\rm{OVII}}$)  $:16.15(16.0-16.3) $. {Although {F15} did not report OVII lines,} \cite{Faerman2017} calculated the median of ratio of N$_{\rm OVII}$ and N$_{\rm OVIII}$ from {G12} observations. {Using this ratio and re-calculated value of N$_{\rm OVII}$ from {F15}, they} calculate log(N$_{\rm OVIII}$) which ranges from $15.3$ to $15.7$ with a median value of $15.5$. The {grey shaded regions} in Figures \ref{OVIII_IT}, \ref{OVIII_PP} and \ref{OVIII_profile} show the ranges of N$_{\rm OVII}$ and N$_{\rm OVIII}$ of {F15} as re-calculated by \cite{Faerman2017}. {Note} that these re-derived ranges of N$_{\rm OVIII}$ indicate $1$-$\sigma$ uncertainty around the median value.

\subsection{Constraints}
We have {tabulated} the {constraints} on $\eta$  for different models in Table \ref{table}. It is evident from Figures \ref{OVIII_IT}, \ref{OVIII_PP} and \ref{OVIII_profile} N$_{\rm{OVIII}}$ can help in putting upper limits on $\eta$. 
 For IT model, we find that { $\eta \le 5$ if one consider only CI with no temperature fluctuations. The inclusion of PI  slightly changes this constraint to $\eta \le 6$. We can also clearly see that introducing log-normal fluctuations makes these constraints even stronger ($\eta \le 1$) (Figure \ref{OVIII_IT}). Interestingly, we find that the models with varying $\eta$ with radius can be accommodated within
observational limits by suitably decreasing $\sigma_{\rm ln T}$ (Figure \ref{OVIII_profile}). For example, in the Figure \ref{OVIII_profile}, the model of increasing $\eta$ with radius for $\sigma_{\rm ln T}=0$ is fairly close to the bottom range of {F15}'s data, and for the opposite case ($\eta=1/x$), decreasing $\sigma_{\rm ln T}$ sufficiently can put in the ballpark of {F15}'s data. } 
}

A larger upper-limit ($\eta \le 8$) is allowed for the case of PP model with T$_{\rm bc1}$ if one  considers of observations of {F15}. In general, we find that the inclusion of temperature fluctuation relaxes these limits in all the cases by allowing larger value of $\eta$ ($\le 10$) (Figure \ref{OVIII_PP}). Furthermore, spatial variation of $\eta$
(Figure \ref{OVIII_profile}) can be allowed within the observational limits by increasing temperature fluctuations and using suitable mass range in PP model. In particular, models with declining ratio of CR to thermal pressure with radius (as hinted in the simulations of, e.g., \cite{Butsky2018}) predict OVIII column densities within observational limits. Note that, in the PP case the requirement for $\sigma_{\ln T}$ for varying $\eta$ models runs opposite to that in IT, where varying $\eta$ models require smaller $\sigma_{\rm ln T}$ in order to be viable. 

{To see the effect of normalization {for the cases $\eta=Ax$ or $=A/x$,} we have increased the {value of} $A$,  which increases the CR pressure. This {leads to a decrease in} the density or temperature or both, and {consequently} decreases the OVIII column density. Therefore, all the plots in Figure \ref{OVIII_profile} will shift towards left with the increment of the normalization. That {implies that} $\eta=A\times x$ cases are not allowed for both IT and PP models even with the inclusion of photoionization and temperature fluctuations. However, in the case of the PP model, for $\eta=A/x$ with A$\ge1$, all the cases do not satisfy the observational constraints except for the case with boundary temperature $1.8\times10^6$ K, {for} $A\le10$. On the other hand, in the case of IT, for the $\eta=A/x$ profiles, the OVIII column densities are within the observational constraints for the cases $A\le8$. {That implies these models allow the central region to have large CR pressure even $\eta=200$, but not in the outer halo. This result differs from the findings by previous simulations which shows CR dominated outer halo \citep{Butsky2018}.} As normalization increment shifts all the plots towards left, the inclusion of temperature fluctuations are ruled out for the models with $\eta=A/x$ profiles except for a very small range {with high CGM mass} in the PP model with boundary temperature $1.8\times10^6$ K and $A\le2$.} 


\section{Discussion}
{Our constraints allow larger values of $\eta$ for most of the cases in comparison to the isentropic model by \cite{Faerman2019,Faerman2021}. In their recent work \citep{Faerman2021}, they have considered three cases for their model : 1) with only thermal pressure i.e, $\alpha=(P_{nth}/P_{th}) +1.0 =((P_{CR}+P_{B})/P_{th}) + 1.0=1.1$,  2) with standard case as considered in \cite{Faerman2019} i.e, $\alpha=2.1$  and 3) with significant non-thermal pressure where $\alpha=2.9$. Note that the {the above mentioned values of} $\alpha$ are the values at the outer boundary and the profiles of $\alpha$  will follow a declining pattern as radius decreases (see blue curve in the right panel of Figure 2 by \cite{Faerman2019}). With {our} definition of $\eta$, along with the magnetic field {value} used by us, we can convert these $\alpha$ values to $\eta \sim 0, 0.6, 1.4$, respectively, for their three cases. The thermal case matches N$_{\rm OVIII}$ observations up to CGM mass of $10^{11} \rm M_{\rm solar}$, and for the standard case, the value of N$_{\rm OVIII}$ is {comparable} to the observed value. However, for the significant non-thermal case, the OVIII value is lower than the observations by a factor of $\sim 10$ for a CGM mass of $10^{11} \rm M_{\rm solar}$. Therefore, one can say that $\eta<0.6$ is allowed for the isentropic model. For PP model, we get upper limit of $\eta \sim 0.5$ for two cases which are therefore in agreement with the constraint from the isentropic model. However, for most of our models, {the upper limits derived in the present work are larger} than the isentropic model.}

It should be noted that the limits of $\eta$ derived here {translate to} constraints on cosmic ray pressure in the hot diffuse CGM, {and one may wonder about the CR pressure in the cold gas}. CR pressure scales with density ($\rm P_{\rm CR} \propto \rho^{\gamma_{c,eff}}$, where $\gamma_{c,eff}$ is the effective adiabatic index of the CR) of the gas, {where the adiabatic index depends} on the transport mechanism \citep{Butsky2020}, the limits of CR pressure in the cold gas would be different from what we have found for the hot gas. If CRs are strongly coupled to the gas, {then} in the limit of slow CR transport, {\it i.e}, if the only CR transport mechanism is advection,  $\gamma_{c,eff}=\gamma_{c}=4/3$. In this case, CR pressure can be higher in cooler, denser gas than in the hot gas. However, in the limit of efficient CR transport i.e, if there are CR diffusion, streaming along with advection, the CR pressure will be redistributed from high density, cold region to diffuse, hot region of CGM which will lower the $\gamma_{c,eff}$. In this limit, $\gamma_{c,eff}\rightarrow0$, {and} the limits of CR pressure in the cold phase are nearly equal to the CR pressure in hot phase. 
{In the context of a single phase temperature with temperature gradient, if one considers  CR pressure in the hot phase to {scale as} $\rho_{g}^{\gamma_{c,eff}}$ then $\eta$ is proportional to $\rho_{g}^{\gamma_{c,eff}-(5/3)}$, as adiabatic index for gas is $5/3$.  Then $\eta$ will be proportional to $\rho_{g}^{-1/3}$ and $\rho_{g}^{-5/3}$ respectively depending on slow and effective transport mechanism. This implies that $\eta$ is smaller in denser gas and hence mimic a rising profile of $\eta$ in case of the single hot phase gas, which we have shown in the present paper.}

{However, these scaling relations are for adiabatic situations, {which is not the case for PP model which {involves} energy loss}. In the context of cold gas that may have condensed out of the hot gas, it is possible for the cold gas to contain a significant amount of CR pressure, with a large value of $\eta$ for cold gas. Simulations by \cite{Butsky2020} (their figure 10) showed that in the limit of slow cosmic ray transport, the value of $\eta$ is quite different for hot and cold gas. In the limit of efficient transport, cosmic ray pressure is decoupled from the gas, and $\eta$ has similar values in hot and cold gas. {If we recall} the observation by \cite{Werk2014}, that the cold phase of CGM nearly follows the hot gas density profile {then for small} density contrast, the {CR pressure} in the cold phase and hot phase {may} not {differ much} even with different transport mechanisms. {But even with small density contrast, thermal pressure of cold gas will be smaller than hot phase by two order of magnitude due to the temperature difference, which allows {for} a larger upper limit on $\eta$.} However, packing more CR in cold gas will not have any effect on the constraints derived here. In fact, this may be one way to circumvent the constraints on $\eta$ described here.}

We have not included any disc component in the present work. A recent work by \cite{Kaaret2020} took into account an empirical disc density profile motivated by the measured molecular profile of MW along with the halo component. They pointed out that the high density disc model can well fit the MW's soft X-ray emission, whereas it under predicts the absorption columns. Their conclusion is that the dominant contribution of X-ray absorption comes from the halo component due to large path length. Also this difference in contribution comes from the fact that the absorption is proportional to density whereas emission measure is proportional to density square.  Therefore, considering only the halo component and making the comparison with the absorption column is justified.\\

{Note that the limits derived here are sensitive to some of the parameters such as the magnetic field and metallicity. Increase (decrease) in the magnetic pressure would increase (decrease) the total non-thermal pressure, which {would} decrease (increase) the limit on $\eta$. However, given the uncertainty in the magnetic field strength in the CGM,  equipartition of magnetic energy with thermal energy is a {reasonable assumption}. For a collisionally ionized plasma, {which} we assume for the CGM gas, the column density is proportional to the metallicity ($Z$) in the case of IT model, whereas it {is} proportional to $Z^{0.3}$ for the PP model (see eq 17 of \cite{Voit2019}). In the photoionized case, the dependence {on} metallicity will {be stronger since} the ion abundance would also depend on the metallicity.}
{{Also note} that the limits derived here refer to CR population spread extensively in the CGM. Localized but significant CR population may elude the above limits. However, we also note that most of the contribution of the limits come from the inner region, where OVIII exists otherwise and is suppressed due to CR. Therefore, our limits also pertain to localized CR populations in the inner regions. }

\section{Summary}
\label{summary}
We have studied two analytical models of CGM (Isothermal and Precipitation model) in hydrostatic equillibrium {in light of} the observations of N$_{\rm{OVIII}}$ {in order to} constrain the CR population in the Milky Way CGM. {We found that} 
$\eta \le 10$ 
 in the case of no photoionization and including temperature fluctuations {in the PP model whereas IT model allows $\eta \le 6$ with inclusion of PI}. 
However, it should be noted that IT is an extremely simplistic model and the inclusion of non-thermal component leads to a very low density. The limits from IT {may therefore be only of academic interest}.

{Combined with the limits derived from $\gamma$-ray background \cite{Jana2020}, our results make it difficult for a significant 
CR population in our Milky Way Galaxy.} 
{However, models in which the ratio of CR to thermal pressure varies with radius (preferably declining with radius) 
can be accommodated within observational constraints, if they include suitable temperature fluctuation (larger $\sigma_{\rm ln T}$ for PP and lower for IT), {although making allowances for large value of $\eta$ only in the central region}.
}

\bibliography{reference.bib}{}
\bibliographystyle{aasjournal}

\end{document}